\documentclass[manuscript]{aastex}

\usepackage{tikz}
\usepackage{amsmath}

\usepackage{graphicx}

\usepackage{pdflscape}

\slugcomment{\textbf{Astronomical Journal, Revised 2019 March 11}}

\shorttitle{3200}
\shortauthors{Jewitt}

\begin{document}

\title{High Resolution Thermal Infrared Imaging of 3200 Phaethon}


\author{David Jewitt$^{1,2}$, Daniel Asmus$^{3,4}$, Bin Yang$^4$
and Jing Li$^{1}$ 
} 

\affil{$^1$Department of Earth, Planetary and Space Sciences,
UCLA, 
595 Charles Young Drive East, 
Los Angeles, CA 90095-1567\\
$^2$Department of Physics and Astronomy,
UCLA, 430 Portola Plaza, Box 951547,
Los Angeles, CA 90095-1547\\
$^3$ Department of Physics and Astronomy, University of Southampton, Southampton
SO17 1BJ, United Kingdom \\
$^4$ European Southern Observatory (ESO), Alonso de C\'ordova 3107, Vitacura, Santiago, Chile \\
}

\email{jewitt@ucla.edu}

\begin{abstract}
We present thermal infrared observations of the active asteroid (and Geminid meteoroid stream parent) 3200 Phaethon  using the Very Large Telescope.  The images,  at 10.7 $\mu$m wavelength, were taken with Phaethon at its closest approach to Earth (separation 0.07 AU) in 2017 December, at  a linear resolution of about 14 km.    We probe the Hill sphere (of radius $\sim$66 km) for trapped dust and macroscopic bodies, finding neither, and we set limits to the presence of unbound dust.   The derived limits to the optical depth of dust near Phaethon depend somewhat on the assumed geometry, but are of order 10$^{-5}$.   The upper limit to the rate of loss of mass in dust is $\lesssim$14 kg s$^{-1}$.  This is $\sim$50 times smaller than the rate needed to sustain the Geminid meteoroid stream in steady state.  The observations thus show that the production of the Geminids does not proceed in steady state.

\end{abstract}


\section{INTRODUCTION}

The Geminid meteoroid stream is a massive complex of sub-millimeter to centimeter (and maybe decimeter) sized solid particles that have been released from their parent body within the last $\sim$10$^3$ yr (Williams and Wu 1993, Ryabova 1999, Beech 2002).   Their reported source is the near-Earth object  3200 Phaethon (hereafter just ``Phaethon''), a B-type (optically blue) body (Bus and Binzel 2002) about 5 km (Hanu{\v{s}} et al.~2016) or 6 km (Taylor et al.~2019) in diameter, with an orbit that is strongly decoupled from Jupiter.  With orbital semimajor axis $a$ = 1.271 AU, eccentricity $e$ = 0.890, and inclination $i$ = 22.2\degr~(orbital elements taken from the JPL Horizons web site at \url{https://ssd.jpl.nasa.gov/horizons.cgi}) the resulting aphelion of Phaethon ($Q$ = 2.40 AU) lies far inside  Jupiter's 5.2 AU radius orbit.  In turn, the Tisserand parameter with respect to Jupiter, $T_J$ = 4.509, lies far above the dividing line separating comets ($T_J \le$ 3) from asteroids ($T_J >$ 3).  Phaethon is thus one of the few asteroidal (as opposed to cometary) stream parents (Kasuga and Jewitt 2019).

Most attempts to identify on-going mass-loss from Phaethon have failed (e.g.~Hsieh and Jewitt 2005, and references therein).  However, activity has been detected close to   perihelion (at distance $q$ = 0.14 AU), first indirectly through an excess brightening that cannot be explained through geometric effects on a body of constant cross-section (Jewitt and Li 2010, Li and Jewitt 2013)  and then directly, through the imaging of a weak dust tail (Jewitt et al.~2013, Hui and Li 2017).  The sub-solar surface temperature of Phaethon at perihelion is $T_{SS} \sim$ 10$^3$ K, leading to the suggestion that the observed mass loss might be a product of thermal fracture and/or desiccation stresses induced in originally hydrated minerals (Jewitt and Li 2010).  Both the position angle and the sudden emergence of the tail  indicate that  the particles released near perihelion are small enough to be strongly accelerated by radiation pressure, with a nominal (although very poorly determined) size $\sim$1 $\mu$m (Jewitt et al.~2013).  Such tiny  particles are distinct from the millimeter-sized Geminids.  Indeed, the  perihelion mass-loss rate inferred from optical observations is only $dM/dt \sim 3$ kg s$^{-1}$, which is orders of magnitude too small to supply the Geminid stream within its $\sim$10$^3$ yr lifetime  (Jewitt et al.~2010).

The absence of a clear mechanism for the production of  the Geminids is a core problem, and is the primary motivation for the present work.  We are interested in the possibility that, far from perihelion, processes other than thermal fracture and desiccation might operate and yet have so far escaped detection.   A secondary motivation is the desire to assess the near-Phaethon dust and debris environment, as a precursor to the planned flyby of the DESTINY+ spacecraft (Arai et al.~2018).  The close approach of Phaethon to the Earth (minimum separation 0.069 AU on UT 2017 December 17) provided an ideal opportunity to search for evidence of mass loss near 1 AU.    A parallel observation was undertaken at optical wavelengths using data from the Hubble Space Telescope (HST; Jewitt et al.~2018a).  HST offers a wide field of view (up to 162\arcsec$\times$162\arcsec) but, because of saturation, scattered light from the main body, and unavoidable trailing (caused by the inability of HST to track at an accelerating non-sidereal rate), the HST data do not  probe distances as close to Phaethon as achieved in the present work.   Independent HST observations by another team, again  near closest approach, were taken to search for boulders distant from Phaethon  (Ye et al.~2018).  However, the telescope in their measurements was pointed away from Phaethon  and so the results by  Ye et al.~cannot be directly compared either to Jewitt et al.~(2018a) or to the present work. 

In this paper, we present thermal infrared observations taken at closest approach in search of near-nucleus emission from solid material.

\section{OBSERVATIONS}
\label{observations}

We observed Phaethon using the  8 meter diameter Very Large Telescope (VLT-U3)  at the European Southern Observatory, Paranal, Chile.  We employed the upgraded VLT mounted Spectrometer and Imager for the Mid-infrared (VISIR), which houses a 1024$\times$1024 pixel Raytheon array with a scale 0.045\arcsec~pixel$^{-1}$ (Lagage et al.~2004, Kaufl et al.~2015, Kerber et al.~2016).  VISIR operates in the 8 $\le \lambda \le$ 13 $\mu$m and 16.5 $\le \lambda \le$ 24.5 $\mu$m wavelength atmospheric transmission windows.

Observations were attempted on both UT 2017 December 17 and 18.  Conditions on the first night were good, with sub-arcsecond optical seeing and no visible clouds while on the  second night clouds were present and the seeing was poor and variable.  Accordingly, in the remainder of the paper we consider only the observations from UT 2017 December 17, which were taken using the  J8.9 ($8.7 \pm 0.37\,\mu$m) and B10.7 ($10.65 \pm 0.68\,\mu$m) filters with total on-source exposure times of 16 and 30\,min, respectively.   Photometric calibration was obtained from standard star HD\,2436 observed at similar airmass ($\sim$1.7) with 120 s exposures per filter, selected from Cohen et al.~(1999).  

Phaethon's heliocentric distance during the observations on UT 2017 December 17 was $r_H$ = 1.010 AU, geocentric distance $\Delta$ = 0.069 AU, and phase angle $\alpha$ = 66.06\degr.   The projected directions of the  anti-solar vector and the negative velocity vector were $\theta_{-\odot}$ = 73.37\degr~and $\theta_{-V}$ = 74.40\degr, respectively, while the Earth was located  1.23\degr~below the plane of Phaethon's orbit.  One consequence of the very small geocentric distance was a highly favorable image scale of only $s$ = 50 km arcsecond$^{-1}$, an order of magnitude smaller than typically achieved in ground-based observations of small solar system bodies at distances $\sim$1 AU.  A less desirable consequence was the rapid angular motion of about 0.6\arcsec~s$^{-1}$ relative to the sidereal background.  The field stabilization of the VLT guiding system used normally for VISIR observations could not simultaneously accommodate such rapid motion and the fast chopping  ($>1\,$Hz) of the secondary mirror as required for observing in the mid-infrared.   Therefore, the observations were performed using ``open-loop tracking'' in which the telescope followed the ephemeris motion of the target with respect to the celestial background but without the use of a guide star.
To compensate for the resulting image drifts and instabilities in the point spread function (PSF), we used the burst mode of VISIR to save every exposure of $\sim 20\,$ms length.
In addition, the canonical chopping and nodding technique was used for background subtraction, with a chop throw of 8 arcsec at a frequency $\sim4$\,Hz, and perpendicular nodding every $\sim 1$\,min.

\begin{figure}
\epsscale{0.95}
\plotone{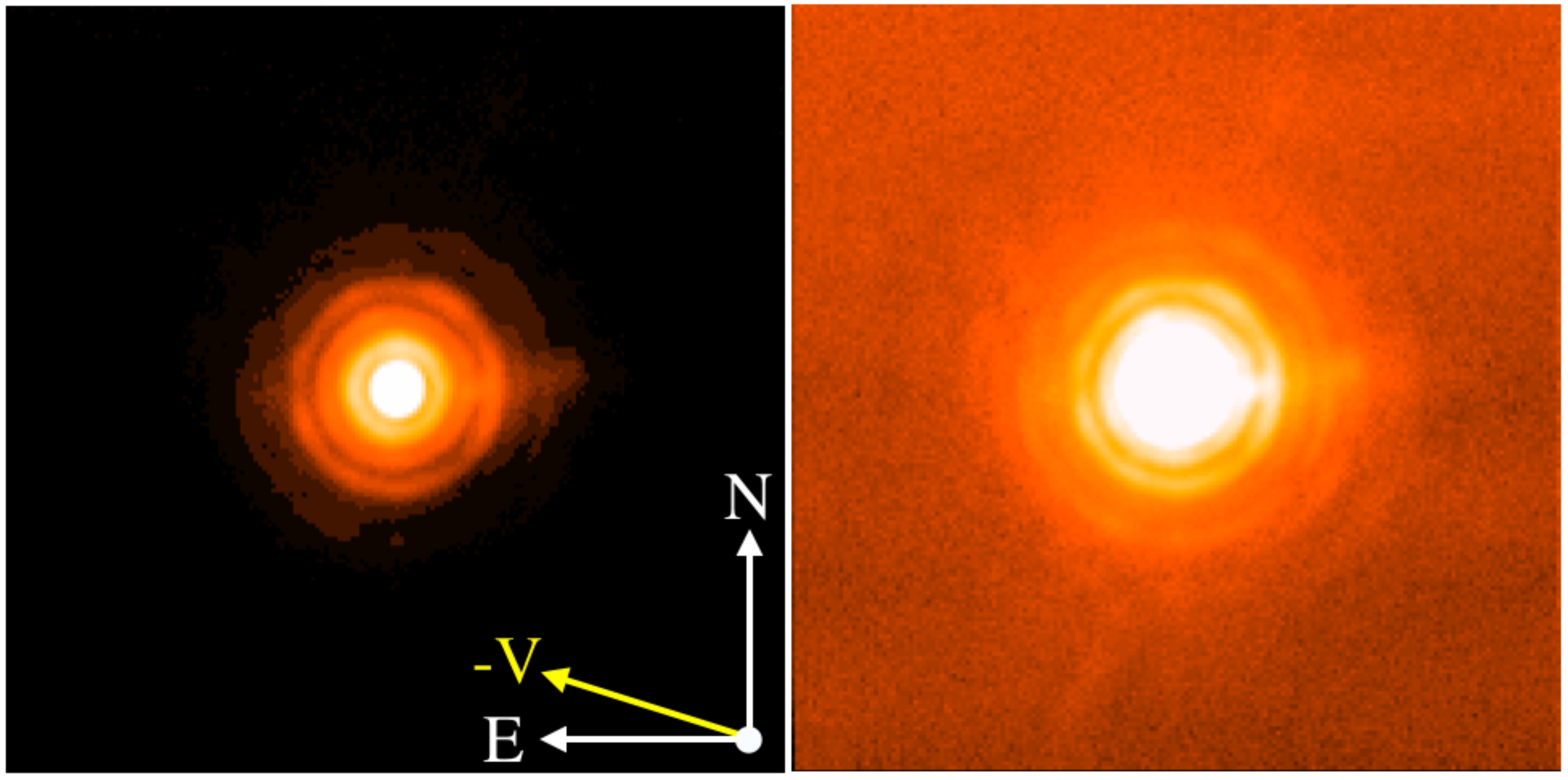}
\caption{(left:) Median-combined 10.7 $\mu$m  image of Phaethon scaled from 0 to 8000 counts pixel$^{-1}$. (right:) Same image but scaled from 0 to 800 counts pixel$^{-1}$ to emphasize fainter structures.  Each panel is 7\arcsec$\times$7\arcsec.  The cardinal directions and the direction of the negative heliocentric velocity vector ($-V$) are marked.  The anti-solar direction is the same as $-V$.
\label{image}}
\end{figure}

\begin{figure}
\epsscale{0.95}
\plotone{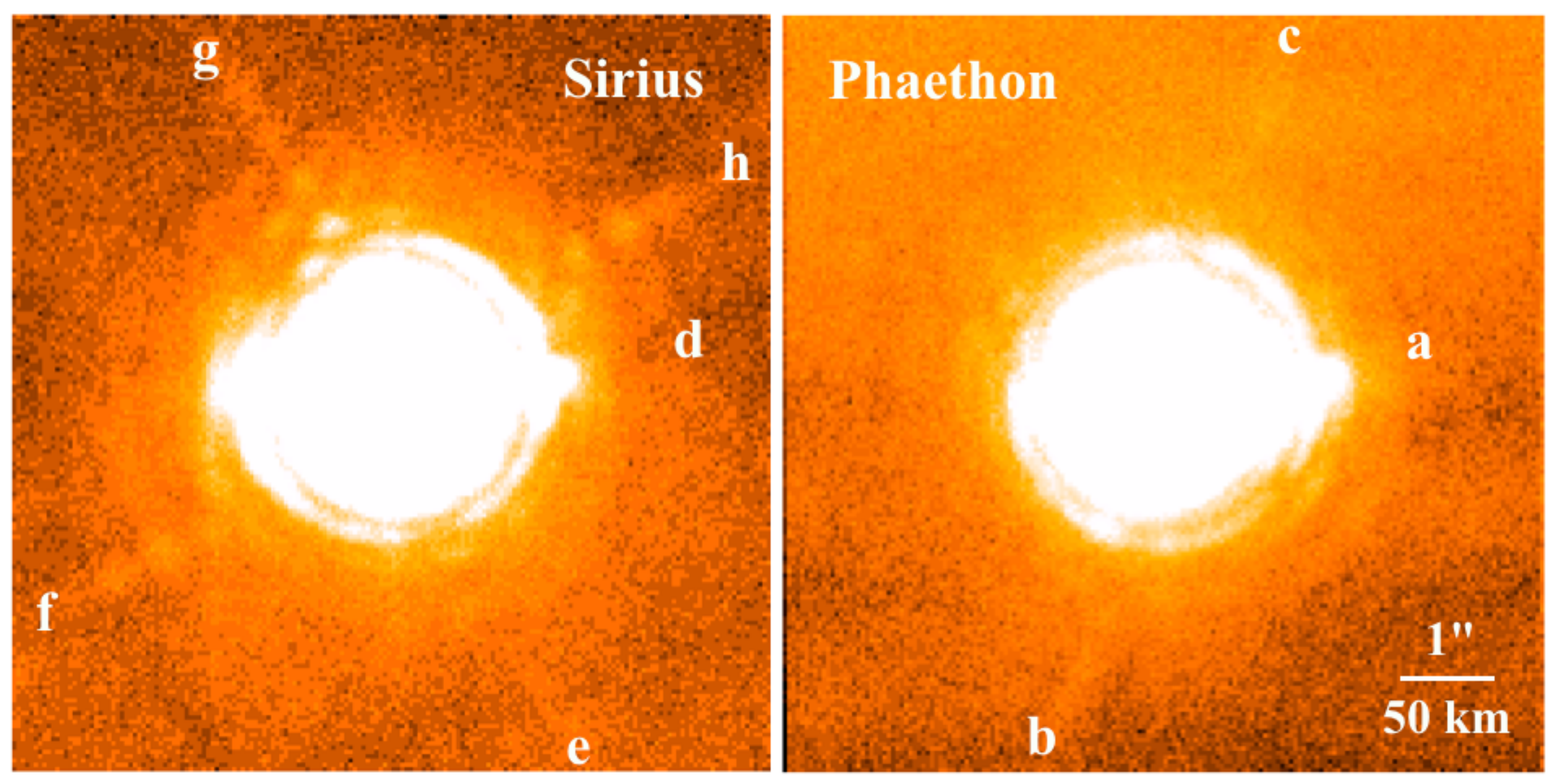}
\caption{(left:) Image of bright star Sirius taken to examine the spider diffraction and spurious light field in VISIR.  The image was obtained in ``pupil tracking'' mode, causing the spider pattern to remain at a fixed position angle.   (right:) Phaethon at a similar stretch, showing similar structure but with the suppression of the spider diffraction pattern due to image rotation in ``field tracking'' mode.    Both images have North to the top, East to the left.  A 1\arcsec~scale bar is shown.
\label{star_comp}}
\end{figure}

Data reduction employed a custom-written \textsc{python} pipeline (Asmus, in prep) which performs  chop and nod subtraction of all individual burst exposures before aligning the latter using Gaussian centroids fitted to the central source.  The final image, formed from a combination of all the images of Phaethon, is shown in Figure (\ref{image}) with two stretches to emphasize the core (left panel) and wing (right panel) portions of the image.  Prominent Airy rings in the figure are testament to the quality of the VLT and the data.  The image of Phaethon is compared with that of the bright star, Sirius, in Figure (\ref{star_comp}).  Sirius was observed as part of programme 098.C-0050 (PI: Sterzik) on UT 2016 December 06 in the same observing mode (B10.7, burst mode) and is used here as a high signal-to-noise PSF reference  owing to its high brightness.  Feature ``a'' in the Phaethon image is replicated at the same position angle in the Sirius image (labelled ``d'') and is thus unrelated to the asteroid.  The stellar image was obtained using ``field tracking'' mode, in which the position angle of the spider diffraction pattern (features ``e'', ``f'', ``g'', and ``h'' in the Figure) rotates and the sky background is fixed.  The spider pattern in the Sirius data is more prominent because of the brightness of the star, and because of the short duration of the measurement (2 minutes) during which time rotation of the pattern is minimal.   Note that the opposing VLT spider arms are laterally offset with respect to each other, so that they do not produce the usual cross-shaped diffraction pattern.  The spider diffraction spikes in the longer duration Phaethon image, while still visible, are much fainter than in the Sirius image.  Peculiarly, only two of the four arms (marked ``b'' and ``c'') are obvious in Figure (\ref{star_comp}).  We checked the sequence of images used to create the Phaethon composite to confirm that all four arms are present in the original data, and that two of them largely vanish in the computation of the median composite. 

The full width at half maximum of the Phaethon image is $\theta_{1/2}$ = 0.31$\pm$0.01\arcsec, very close to the diffraction limit of the system ($\theta_D$ = 1.03$\lambda/D$ = 0.28\arcsec, where $D$ = 8 m is the telescope diameter).  The  difference, $\theta_{1/2} - \theta_D$ = 0.03$\pm$0.01\arcsec, teeters on the edge of statistical significance and is consistent with the finite angular diameter of Phaethon (a diameter $D \sim$ 6 km at $\Delta$ = 0.069 AU subtends angle $\theta_{Ph}$ = 0.13\arcsec~and, approximated as a Gaussian, gives composite width  $(\theta_D^2 + \theta_{Ph}^2)^{1/2} =$ 0.31\arcsec).
 
\begin{figure}
\epsscale{0.80}
\plotone{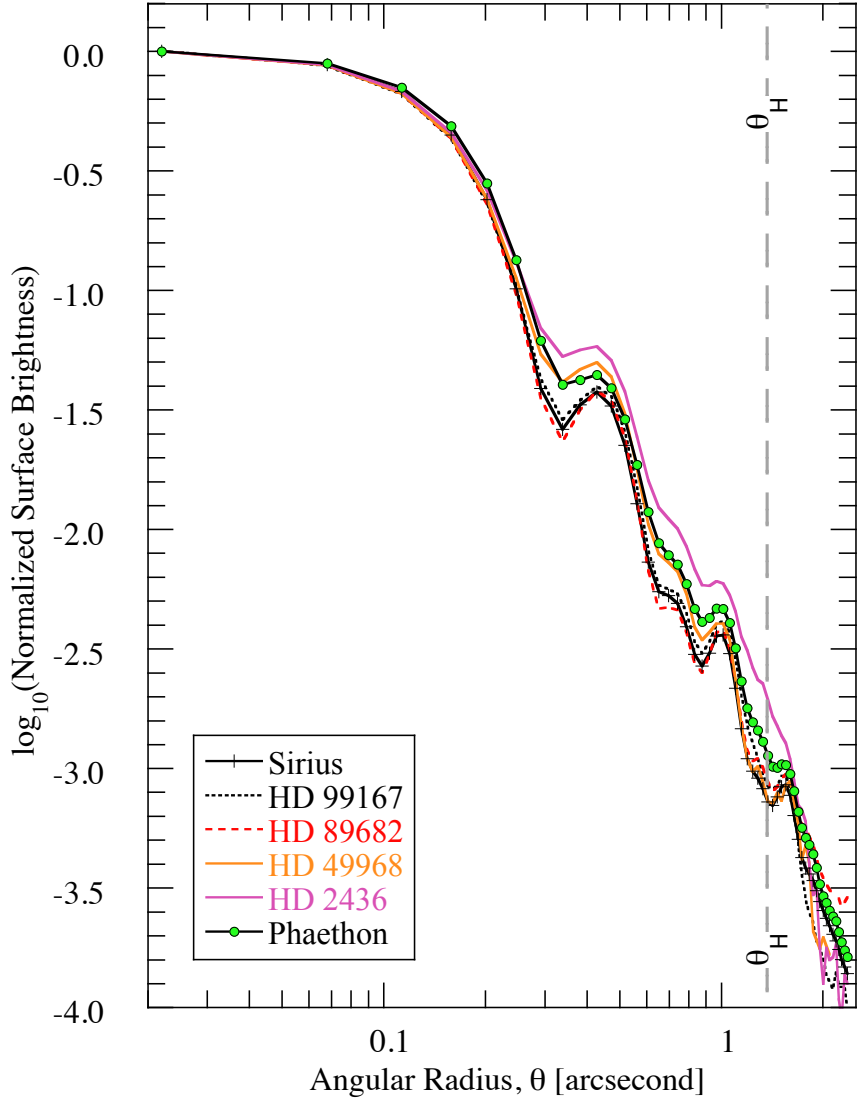}
\caption{Surface brightness profiles of five stars and Phaethon, as labeled, showing the dispersion of the point spread function of VISIR measured with different times, airmasses and signal-to-noise ratios. The widest profile (HD 2436) is thought to be influenced by the lower signal to noise ratio in this object.  The difference between Phaethon and the remaining stars is possibly affected by the different airmasses at which the data were taken. The profiles were all determined in the same way, using concentric circular apertures centered on the optocenter of each image. \label{allstars}}
\end{figure}

\begin{figure}
\epsscale{0.8}
\plotone{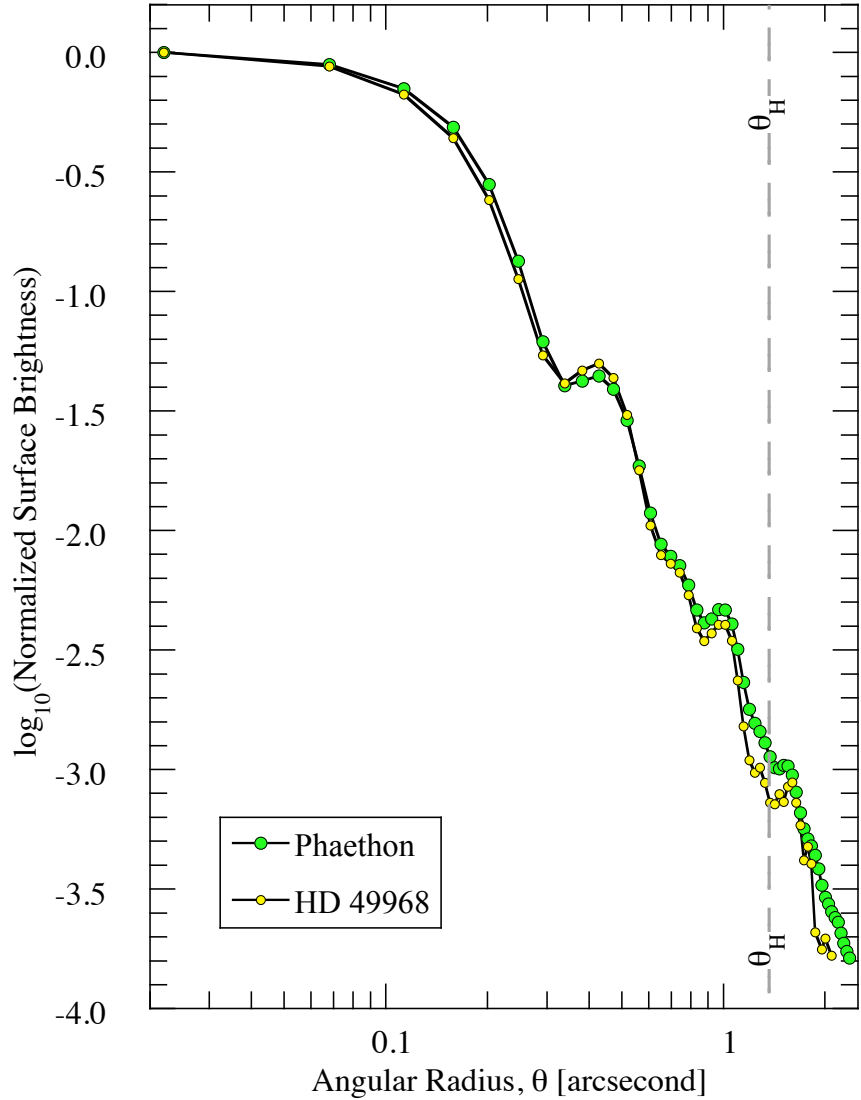}
\caption{Normalized surface brightness profiles of Phaethon (black line, green circles) and star HD 49968 (yellow circles) compared.  The profiles were determined using concentric circular apertures centered on the optocenter of each image.  The profile of Phaethon is scaled such that the peak corresponds to 400 Jy arcsecond$^{-2}$. The angular radius of Phaethon's Hill sphere is marked with a vertical dashed line.
\label{ph_star}}
\end{figure}

Star HD\,2436 was observed on the same night as Phaethon, albeit at higher airmass. In addition, we examined archival images of  other stars obtained in burst-mode with VISIR in order to better characterize the properties of the PSF.  We found that the surface brightness profiles of the stars are not all the same.  The bright stars HD 89682 (14 Jy) and HD 99167 (16 Jy) were observed as part of programme 60.A-9800(I) for calibration purposes between May 2016 and May 2018.  Together with Sirius (125 Jy), and even though measured on different dates and under different conditions, these three bright stars  have profiles that are  consistent with each other, both in the core and the wings (Figure \ref{allstars}).  However, while the FWHM are the same as in Phaethon, the surface brightnesses of the wings of the stars are smaller than in Phaethon (marked with green circles in Figure \ref{allstars}), while the profile of HD 2436 (5 Jy) has wings brighter than in Phaethon.     To test this apparent signal-to-noise ratio (SNR) dependence, we examined three images formed from subsets of the Phaethon data.  These images show clear speckle patterns in the inner Airy rings owing to incompletely averaged short-term fluctuations in the atmosphere  and have brighter wings than the full Phaethon integration, confirming an effect dependent on SNR.  The elevation of the telescope constitutes a second concern.  The Phaethon data were taken at airmass $\chi \sim$ 1.7, while the stars were observed at $\chi \lesssim$ 1.3.  This airmass difference introduces a potential bias both because the seeing varies with airmass and because, while the telescope focus is normally updated using images from the guide camera, the lack of a guide star rendered this impossible on Phaethon.  Instead, we set the focus before observing Phaethon and held it fixed throughout the observation.  As a result, it is possible that the Phaethon images are slightly less well focused than the archival stars. In this regard, we note that the profile of star HD 49968 (which was observed at similar airmass and seeing to Phaethon, and which has integrated flux density of 5 Jy) is consistent with that of Phaethon.   Normalized surface brightness profiles from Phaethon and HD 49968  are shown in Figure (\ref{ph_star}), where they are seen to be consistent.  

We conservatively conclude that the profile of Phaethon provides no evidence for extended emission.

\section{DISCUSSION}

\subsection{Temperature}

The  measured flux densities of Phaethon on UT 2017 December 17 were  $S_{\nu}(8.9) = 39.8 \pm 4.0$\,Jy and $S_{\nu}(10.7) = 54.9\pm5.5 $\,Jy in J8.9 and B10.7 filters, respectively, where the uncertainties are dominated by the systematic flux calibration uncertainty of the standard star.   The ratio $S_{\nu}(8.9)/S_{\nu}(10.7)$ = 0.73$\pm$0.10 is consistent with the ratio of flux densities expected from a blackbody having temperature $T = 316_{-45}^{+62}$ K.  For comparison, the equilibrium temperature of an isothermal, blackbody sphere located at the same heliocentric distance as Phaethon, $r_H$ = 1.01 AU, would be $T_{BB}$ =  277 K, while  a flat, blackbody plane oriented perpendicular to the Sun - Phaethon line would have $T_{BB}$ = 391 K.   The measured temperature is thus intermediate between these extremes, consistent with the fact that the body of Phaethon is neither a sphere nor a plane and neither is its surface likely to be a perfect blackbody.  In addition, Phaethon was observed from a large phase  angle, $\alpha$ = 66\degr, exposing parts of both the hot dayside and the cold nightside  to view.  We make no attempt to interpret $S_{\nu}(8.9)/S_{\nu}(10.7)$ in terms of the surface thermal thermophysical parameters given both the considerable ($\sim$10\%) systematic uncertainties on this ratio and the large number of unconstrained thermophysical parameters upon which it depends.

\subsection{Surface Brightness Limits: Unbound Coma}
\label{unbound}

Particles ejected faster than the escape speed from the nucleus form a diffuse, unbound coma, as is typical in the active comets.  While no such coma is evident in our Phaethon data,  we are interested to know what limits to coma and mass loss can be placed by this non-detection. 

We define $\tau(r)$ as the optical depth at radius, $r$, in the coma.  Provided $\tau(r) < 1$ we may relate $\tau(r)$ to the surface brightness of the coma, since both are  proportional to the radiating cross-section of dust per square arcsecond on the sky.  To calibrate the latter, we note that the Phaethon nucleus, of cross-section $\pi r_n^2$ = 28 km$^2$, generates a flux density $S_{\nu}(10.7)$ = 55 Jy.

\begin{figure}
\epsscale{0.8}
\plotone{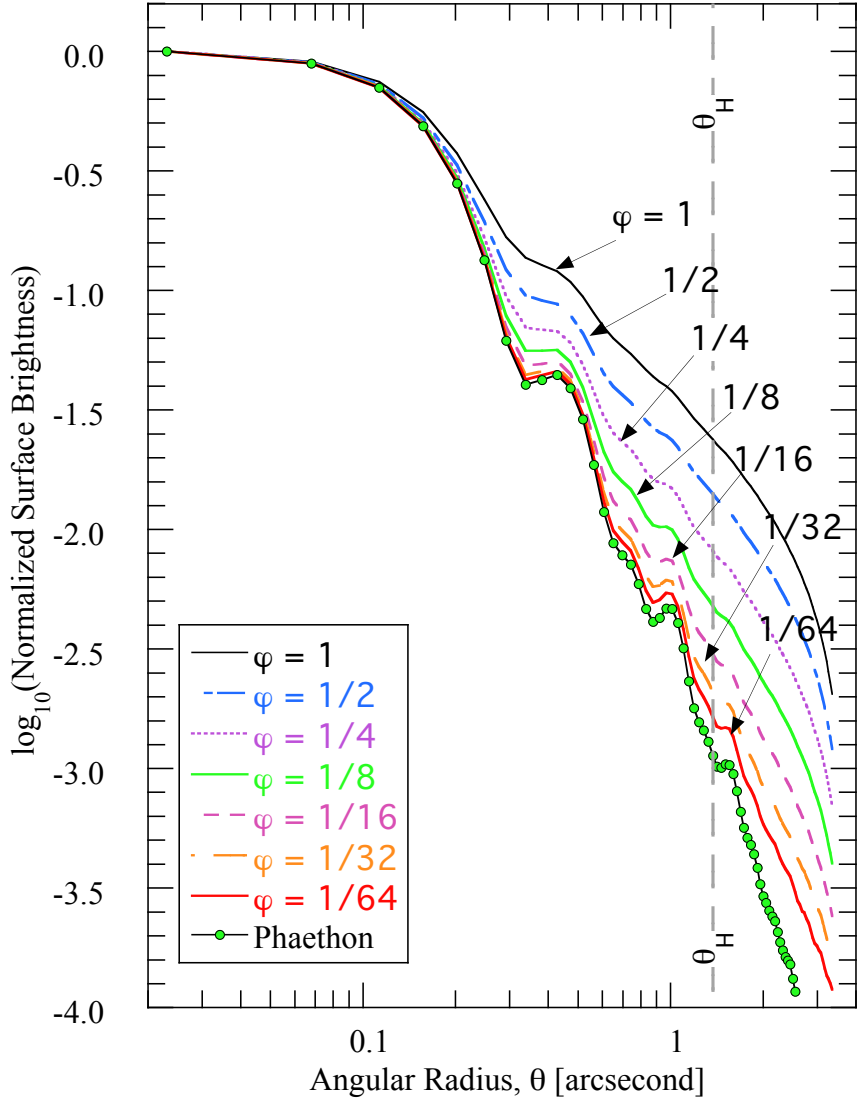}
\caption{Normalized surface brightness profiles of Phaethon with varying amounts of unbound dust, as represented by the $\varphi$ parameter ($\varphi$ = 0 indicates no dust, $\varphi$ = 1 indicates Hill sphere dust with a total cross-section equal to that of the main body of Phaethon). The measured surface brightness profile is shown in black with individual data points as green circles.  The vertical, dashed line marks the angle subtended by the Hill sphere (c.f.~Equation \ref{hill}).   \label{convolution}}
\end{figure}

We convolved the PSF from HD 49968 with models of a steady-state, isotropic coma in which the surface brightness varies as $\Sigma(\theta) = \Sigma(0) \theta^{-1}$.  The constant $\Sigma(0)$ was chosen to give a fixed fractional brightness for each model, defined by the dimensionless quantity $\varphi$, equal to the ratio of the brightness of the model within $\theta \le$ 2\arcsec~to that of Phaethon within the same angle.  After convolution, normalized surface brightness profiles were determined from the model images using the same parameters as employed on Phaethon (namely, gaussian centering on the image, with background subtraction from an annulus extending from 3.4\arcsec~to 3.8\arcsec).  The models are shown together with the Phaethon profile in Figure (\ref{convolution}).  Note that the downturn of the models at large $\theta$ is forced by the background subtraction used in the photometry.  

Figure (\ref{convolution}) shows the expected systematic flaring of the surface brightness profile as the contribution from the coma increases (large $\varphi$).  To set a conservative limit to the possible contribution of the coma, we note that the difference between the Phaethon and model profiles at $\theta$ = 2\arcsec~is comparable to the difference between the Phaethon and star profiles in Figure (\ref{ph_star}) when $\varphi$ = 1/64.  Therefore, we take this as a limit to the fractional contribution of the coma.  With the nucleus cross-section $C_e = \pi r_n^2$ = 28 km$^2$, we find that the cross-section of dust within 2\arcsec~of the nucleus of Phaethon must be $C_d \le$ 0.44 km$^2$ in order to satisfy the measured surface brightness profile.   The resulting coma optical depth is given by $\tau = C_d/(\pi r^2)$, where $r$ = 100 km is the linear distance corresponding to $\theta$ = 2\arcsec.  We find $\tau \le 1.4\times 10^{-5}$.

If the dust grains  are moving radially outward at speed $V_d = dr/dt$ then, to maintain steady-state, fresh cross-section must be injected into the aperture at rate  $dC_d/dt = C_d V_d/r$.  Given that $\tau \ll 1$, we may write the mass of an assemblage of spheres of average radius $\overline{a}$ as $M_d = 4 \rho \overline{a} C_d / 3$.  Differentiating this relation, and substituting for $dC_d/dt$, we obtain

\begin{equation}
\frac{dM}{dt} = \frac{4 \rho \overline{a}  C_d V_d}{3r}.
\label{dmbdt}
\end{equation}

An estimate of the density of Phaethon has been recently proposed  as $\rho \sim  1700\pm500$ kg m$^{-3}$ (Hanus et al.~2018) and, although this estimate is highly model-dependent, we use it here.  Then, we evaluate Equation (\ref{dmbdt}) using $\overline{a}$ = 0.5 mm (the nominal size of the Geminid meteoroids) and $r$ = 10$^5$ m (corresponding to $\theta$ = 2\arcsec) to find $dM/dt \le 5~V_d$ kg s$^{-1}$.  For escape, we require $V_d \ge V_e$, where  $V_e = (8 \pi G \rho/3)^{1/2} r_n$ is the gravitational escape speed.  Substituting, we find $V_e =$ 2.7 m s$^{-1}$.  Therefore, by substitution we find that the  limit to the surface brightness sets an upper limit to the dust mass loss rate at $r_H$ = 1 AU of $dM_d/dt \sim$ 14 kg s$^{-1}$.

\subsection{Surface Brightness Limits:  Bound Coma}
\label{bound}

Slowly-launched dust  might exist in temporarily bound  orbits about the nucleus.  For example, ejected  dust particles could be  trapped into bound orbits by a combination of torques from the aspherical shape of the nucleus, outgassing forces (if present) and radiation forces.  A cocoon of such near-nucleus dust could contribute to a broadening of the PSF without much affecting the more distant wings of the image, far from the core.  We next consider limits to bound dust placed by the surface brightness profile.

The  maximum extent of the region over which Phaethon can exert gravitational control in competition with the Sun is given by the Hill radius, $r_{Hill} = q (m/(3M_{\odot}))^{1/3}$, where $q = a(1-e)$ is the perihelion distance, $m$ is the mass of Phaethon and $M_{\odot}$ is the mass of the Sun.  In terms of the density and radius of Phaethon, $\rho$ (kg m$^{-3}$) and  $r_n$ (m), respectively, we write

\begin{equation}
\theta_{Hill} = \frac{a(1-e)}{\Delta}  \left(\frac{r_n}{r_\odot}\right) \left(\frac{\rho}{3\rho_\odot}\right)^{1/3}
\label{hill}
\end{equation}

\noindent  where $\theta_{Hill}$ is the angle subtended (in radians) by $r_{Hill}$. With $a(1-e)$ = 0.14 AU, $\Delta$ = 0.07 AU, $\rho$ = 1700 kg m$^{-3}$, Phaethon radius $r_n$ = 3 km, solar density $\rho_{\odot}$ = 1300 kg m$^{-3}$ and solar radius $r_{\odot} = 7\times10^5$ km, we obtain $\theta_{Hill}$ = 1.3\arcsec~and  $r_{Hill}$ = 66 km.  The Hill sphere of Phaethon should therefore be resolved in our data, with $2\theta_{Hill}/\theta_{1/2} \sim$ 8 resolution elements across its diameter.

The surface brightness profile presented by a dust-filled Hill sphere depends on the unknown spatial distribution of dust within the sphere.  In the simplest case, with dust distributed at constant number density,  the Hill sphere surface brightness should vary  in proportion to the line of sight path length through the sphere, provided the Hill sphere is optically thin.  Specifically, the surface brightness as a function of the angle from the center, $\theta$, would in this case follow

\begin{equation}
\Sigma(\theta) = \Sigma_0 \left(\frac{\theta_{Hill}^2 - \theta^2}{\theta_{Hill}^2}\right)^{1/2}
\label{three}
\end{equation}

\noindent where $\Sigma_0$ is a constant and  $\theta$ falls in the range $0 \le \theta \le \theta_H$, otherwise $\Sigma(\theta > \theta_H)$ = 0. In practice, the number density inside the Hill sphere is likely to be strongly concentrated towards smaller radii; observations (e.g.~of the irregular satellites of the planets) and models  show that typically only the inner $r_{Hill}/2$ is occupied by long-term stable objects, the more distant ones being prone to escape.  Consequently, the detailed radial dependence of the surface brightness must be considered unknown, except that it is bounded by Equation (\ref{hill}) and likely peaked towards the center (as Equation \ref{three}).

\begin{figure}
\epsscale{0.8}
\plotone{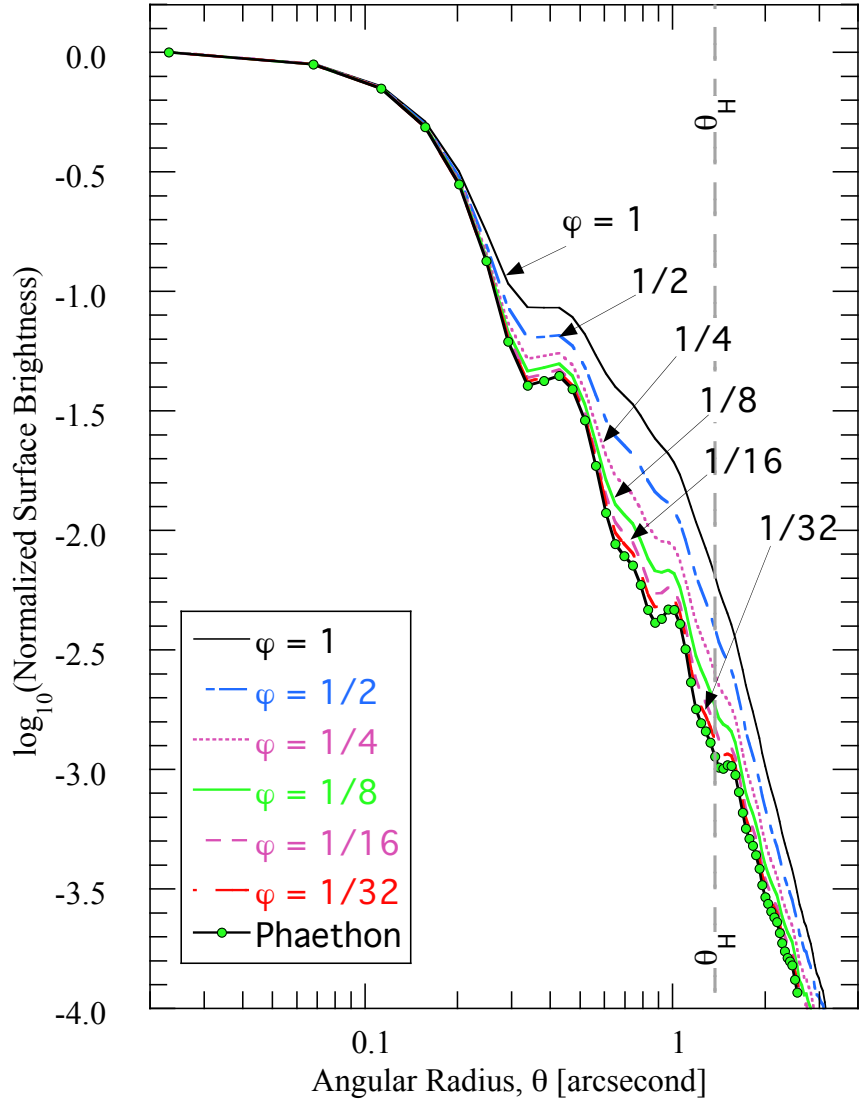}
\caption{Same as Figure (\ref{convolution}) but for bound dust, as represented by the $\varphi$ parameter ($\varphi$ = 0 indicates no dust, $\varphi$ = 1 indicates Hill sphere dust with a total cross-section equal to that of the main body of Phaethon).   \label{hillspheres}}
\end{figure}

With these uncertainties in mind, we  represent the Hill sphere by a Gaussian with FWHM 2$\theta_{Hill}$ = 2.6\arcsec~(standard deviation $\sigma$ = FWHM/2.355 = 1.1\arcsec).  We computed simulated profiles by convolving this Gaussian with the PSF and used the models to obtain a constraint on $\varphi$, defined as the ratio of the signal from the dust in the Hill sphere to the signal from the unresolved nucleus without dust, both within $\theta$ = 2\arcsec.  The convolution kernal was extended to 4$\sigma$ from the Gaussian center and was set equal to zero at larger radii.   Sample profiles computed in this way are shown in Figure (\ref{hillspheres}), for $\varphi$ = 1, 1/2, 1/4...1/32.  As expected, the Figure shows that the model profiles are, for a given value of $\varphi$, less broad than those from the unbound coma model (Figure \ref{convolution}) because of the compact nature of the Hill sphere.  The models show that the FWHM of the image provides a poor measure of the presence of a dusty Hill sphere (the FWHM barely increases  from 0.31\arcsec~at $\varphi$ = 1/32 to 0.32\arcsec~at $\varphi$ = 1), but the profiles are more distinct in the wings, especially beyond the first Airy ring.  Based on Figure \ref{hillspheres}, we take $\varphi$ = 1/32 as a practical upper limit to the fractional cross-section of a  bound dust population.  Proceeding as before, with effective Phaethon cross-section $C_e$ = 28 km$^2$, we find a limit to the dust cross-section in the Hill sphere of $C_e$ = 0.9 km$^2$.  The implied average optical depth is then $\tau = C_e/(\pi r_{Hill}^2)$.  Substituting $r_{Hill}$ = 66 km we find $\tau \le 7\times 10^{-5}$.

Small particles cannot be retained by Phaethon, because of the influence on their motion of radiation pressure.  As an approximate criterion by which to determine the minimum trapped size, we assume that particles can only be held in orbit by Phaethon when the magnitude of the radiation pressure acceleration is small compared to the gravitational attraction to the nucleus.  The acceleration due to radiation pressure is conventionally written $g_{\odot} \beta $, where $\beta$ is a dimensionless factor, $g_{\odot} = G M_{\odot}/r_H^2$ is the local gravitational acceleration to the Sun,  $G$ is the gravitational constant and $M_{\odot} = 2\times 10^{30}$ kg is the mass of the Sun.   At the edge of the Hill sphere, the stability criterion is expressed as

\begin{equation}
\beta  \frac{G M_{\odot}}{r_H^2}  < \frac{G m}{r_{Hill}^2}
\end{equation}

\noindent where $m$ is the mass of Phaethon.  Then

\begin{equation}
\beta   < \frac{m}{M_{\odot}} \left(\frac{r_H} {r_{Hill}}\right)^{2}
\label{beta}
\end{equation}

\noindent which we write as

\begin{equation}
\beta   < \left(\frac{\rho}{\rho_{\odot}}\right) \left(\frac{r_n}{r_{\odot}}\right)^{3} \left(\frac{r_H} {r_{Hill}}\right)^{2}.
\label{beta2}
\end{equation}

\noindent With  radius $r_n$ = 3 km, density $\rho$ = 1700 kg m$^{-3}$ and $r_H = 1.5\times10^{11}$ m, Equation (\ref{beta2}) gives $\beta < 5\times 10^{-4}$.  Finally, $\beta$ is inversely related to the particle size such that, for dielectric spheres,  $\beta \sim a_{\mu}^{-1}$, where $a_{\mu}$ is the particle radius in microns (Bohren and Huffman 1983).  Therefore, we conclude that any particles held in Phaethon's Hill sphere should have $a >$ 2000 $\mu$m (2 mm) and this limit is specific to particles recently launched into the Hill sphere, with $r_H \sim$ 1 AU.   At perihelion, where $r_H = 0.14$ AU, Equation (\ref{beta2}) gives a more severe limit on $\beta$ about 50 times smaller, and on particle radii 50 times larger (i.e.~10 cm).  We thus infer that Phaethon's Hill sphere should be effectively cleaned by radiation pressure of all but the largest bound particles at each perihelion passage.  Only  material released after the last perihelion could still be present.  This inference is consistent with our non-detection of a bound coma at 10 $\mu$m and with the reported absence of  diffuse backscatter in radar data, where the sensitivity is to particles larger than $\gtrsim$cm in size (Taylor et al.~2019). Nevertheless, it is important to search for such material.

\subsection{Surface Brightness Limits: Dust Trail}
\label{trail}

Large dust particles launched barely faster than the escape speed from the nucleus  follow heliocentric orbits close to that of their parent body, forming a narrow ``trail'' when observed in the plane of the sky (e.g.~Reach et al.~2007, Ishiguro et al.~2009).  Such trails are common in the orbits of short-period comets and of the active asteroids, where both the ejection speeds and the trail widths can be incredibly small (e.g.~$<$ 1 m s$^{-1}$ and $<$ 1\arcsec, respectively; Jewitt et al.~2015a).  No such narrow trail, which would be parallel to the $-V$ vector in Figure (\ref{image}), is evident in the Phaethon thermal data, but the structured background makes it difficult to set a uniformly applicable, statistical upper limit to the surface brightness of such a feature.   We simply note that the spider diffraction arms (``b'' and ``c'' in Figure \ref{star_comp}) form a suitable  analogue of a particle trail.  Their peak surface brightness, measured to be 26 mJy arcsecond$^{-2}$ averaged over distances $2.9 \le \theta \le$ 4.0\arcsec, sets a simple and practical upper limit to the surface brightness of any  natural trail.  Scaling from Phaethon, this surface brightness is equivalent to a radiating cross-section 0.014 km$^2$ arcsecond$^{-2}$.  With 1\arcsec~= 50 km, the upper limit to the optical depth in a particle trail is $\tau < 6\times10^{-6}$.

\subsection{Point Sources}
The kilometer-sized asteroid 2005 UD has been reported to show a dynamical association with Phaethon (Ohtsuka et al.~2006) as has, albeit with less certainty, asteroid 1999 YC (Ohtsuka et al.~2008).  The B-type optical color of 2005 UD is similar to that of Phaethon, supporting a physical association (Jewitt and Hsieh 2006), while 1999 YC, with a C-type spectrum (Kasuga and Jewitt 2008) and a more distant orbit, is less obviously related.  Nevertheless, the existence of at least one related kilometer-sized asteroid raises the prospect that Phaethon has fragmented (Kasuga 2009), presumably on a timescale much longer than the $\sim$10$^3$ yr dynamical age of the Geminid stream.  While kilometer-sized fragments in the  near-nucleus space would be immediately obvious in our data, smaller bodies  could linger and yet escape detection.   Accordingly, we sought to set limits to the possible brightness of  point source objects in the VISIR data.

\begin{figure}
\epsscale{0.8}
\plotone{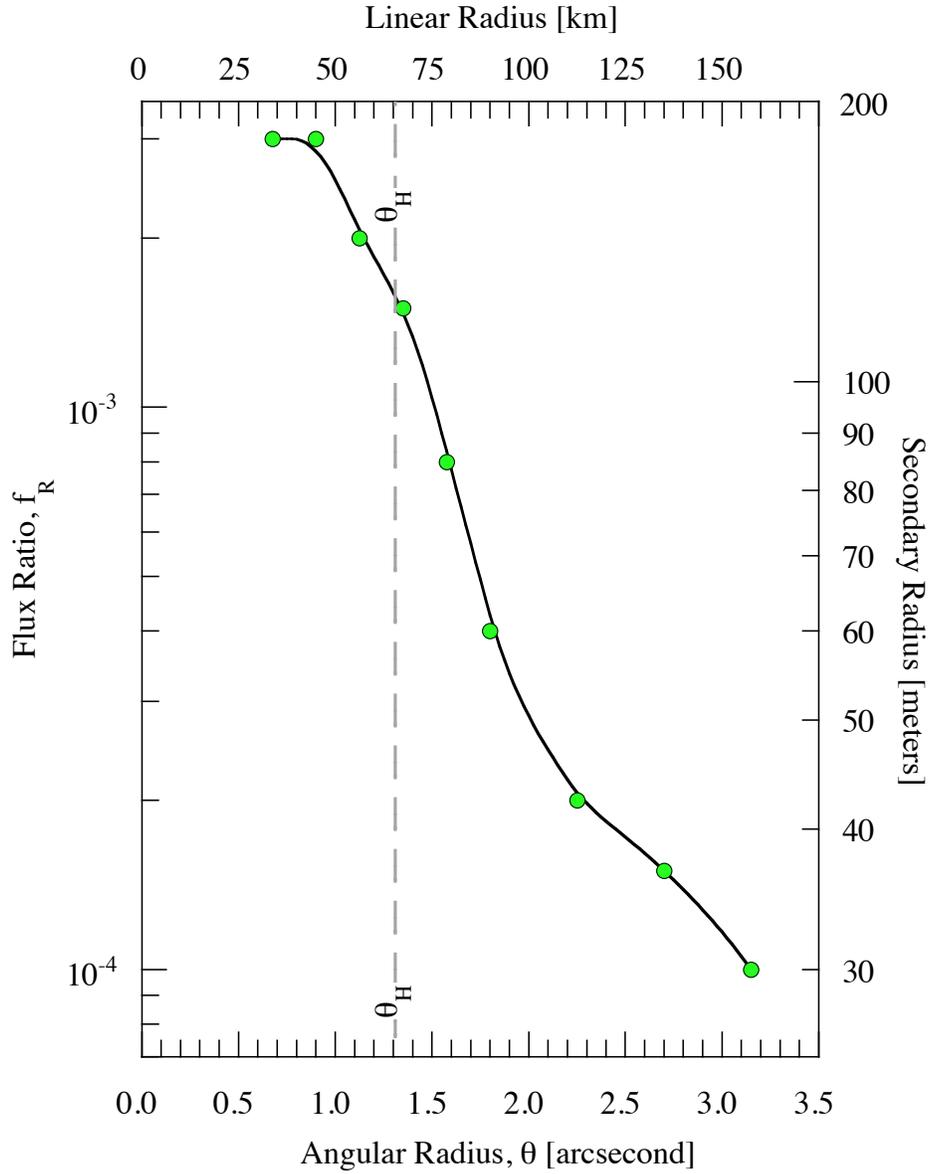}
\caption{Minimum detectable point-source flux  (relative to the Phaethon flux) as a function of angular separation from Phaethon.    Scale on the upper horizontal axis shows the linear scale at the distance of Phaethon.  The right hand vertical axis shows the minimum detectable radius, assuming thermal properties the same as those of Phaethon. The solid black line is a smoothed curve added to guide the eye.  The vertical, dashed line marks the angle subtended by the Hill sphere (c.f.~Equation \ref{hill}).   \label{points}}
\end{figure}

We searched for possible companions by digitally adding to the Phaethon data a set of artificial point source (i.e.~Airy disk) objects with a range of brightnesses and projected distances from Phaethon.  To enhance the detectability of faint sources near the bright image core, we first self-subtracted the images after rotating by 90\degr~about the opto-center of Phaethon.  Our numerical experiments immediately showed that the visibility of companion objects depends not just on the radial distance from the bright core of the Phaethon image, but also on the presence of discrete  brightness structures  within each image.   These residual structures, caused by  diffraction and spurious sources within the optics, remain as a lumpy texture in the difference images when viewed at high contrast.  
To attempt to quantify these spatial variations, we defined the empirical detection limit as occurring  when 50\% of the added artificial companions at a given angular separation were visually detected.  The results are plotted in Figure (\ref{points}), in which the vertical axis shows the point source flux ratio, $f_R$, equal to the ratio of the flux in the added companion  to that of Phaethon.  No useful limits can be placed at angular separations $\theta \lesssim 0.6$\arcsec~because of the brightness of the image core.  From 0.6\arcsec~to the edge of the Hill sphere ($\theta_{Hill}$ = 1.3\arcsec) the point source flux ratio decreases from $f_R \sim 3\times 10^{-3}$ to $f_R \sim  1.5\times 10^{-3}$ because of the fading of the PSF.  Beyond $\theta_{Hill}$ the sensitivity improves less rapidly with increasing separation, reaching $f_R\sim 10^{-4}$ (meaning that 50\% of the objects with $f_R = 10^{-4}$ are detected) by the edge of the image field (at $\theta$ = 4\arcsec).   These large-angle detection limits, which correspond to flux densities $\sim$5 mJy in the B10.7 filter, are broadly in-line with the reported instrument sensitivities (\url{https://www.eso.org/sci/facilities/paranal/instruments/visir/inst.html}) for this filter and integration time, against clean sky.

In order to interpret these limits we scale from Phaethon, assuming that the  flux density  is proportional only to the cross-sectional area of the radiating body. Then,  the limiting radius for possible Phaethon companions, $r_e$ (m), is given by $r_e = 3000 f_R^{0.5}$, and this quantity is shown on the right-hand axis of Figure (\ref{points}).  At the edge of the Hill sphere (separation of 66 km), companions with $r_e =$ 120 m would be individually detected while this limit rises to $\sim$160 m at 0.6\arcsec~(30 km), the inner edge of the useful data.  Outside the Hill sphere, objects down to $\sim$30 m radius would be evident against empty sky in the VISIR data, but are not seen.

\subsection{MECHANISMS}
Independent measurements with the Hubble Space Telescope yield a limit to the optical depth of a Phaethon-associated dust trail, $\tau \le 3\times10^{-9}$ (Jewitt et al.~2018a).  This is three orders of magnitude more stringent than the limit placed above (section \ref{trail}) using VISIR data, reflecting the practical difficulties of mid-infrared observing from a ground-based telescope vs.~observing in the optical from space.   However, the VISIR optical depth limits derived for observations in and near the Hill sphere (sections \ref{unbound} and \ref{bound}) have no counterpart in the optical observations because of near-nucleus saturation, scattering and trailing in the Hubble data.  It should be noted that  optical depths measured at different wavelengths should not be directly compared because of the particle size dependence of the dust radiating efficiency. The optical depth is largely determined by the cumulative cross-section of particles with $a > \lambda$.  For plausible size distributions the effect is modest and, given that we obtained only limits to the optical depth,  moot.

The  mass of the Geminids is $2\times 10^{13} \le M_G \le 7 \times 10^{13}$  kg according to Blaauw (2017) and $10^{13} \le M_G \le 10^{15}$ kg according to Ryabova (2017).  By comparison, the  perihelion mass-loss rate inferred from optical observations is only $dM/dt \sim 3$ kg s$^{-1}$ (Jewitt et al.~2010).  If, as indicated by obervations, the mass loss is sustained for $\sim$1 to 2 days around perihelion, the ejected mass is $\Delta M \sim 5\times 10^5$ kg per orbit.  The time needed to supply the Geminid mass at this rate is $t \sim (M_G/\Delta M) P_o$, where $P_o$ = 1.4 yr is the orbital period of Phaethon.  Even with the minimum Blaauw mass estimate, $M_G = 2\times10^{13}$ kg, this time is $t \sim 6\times10^7$ yr,  orders of magnitude longer than the $\sim$10$^3$ yr dynamical lifetime of the stream.  The mis-match would be even larger  if the higher stream mass estimates of Ryabova (2017) were to be used.  The supply problem is in fact even more acute, because the micron-sized particles released at perihelion are so strongly accelerated by solar radiation pressure that they cannot enter the orbit-hugging Geminid stream.     

What is needed to resupply the Geminid stream mass, $M_G$, in steady-state over the stream lifetime, $\tau_s$, is a source with rate $dM/dt \sim M_G/\tau_s$.  Taking minimum mass estimate, $M_G = 2\times 10^{13}$ kg (Blaauw 2017) and lifetime $\tau_s$ = 10$^3$ years gives $dM/dt \sim $ 700 kg s$^{-1}$, comparable to the mass loss rates from  conspicuously active Jupiter family comets.   We briefly discuss possible mechanisms for mass loss.

\textit{Thermal Fracture and Desiccation Stress}: Peak perihelion temperatures $\sim$10$^3$ K are sufficient to cause thermal fracture of exposed rocks and also to cause desiccation and shrinkage cracking of hydrated minerals, if present (Jewitt and Li 2010).  If even a few percent of the stress energy built up by these processes is converted into kinetic energy, then the resulting particles  leave the surface of a kilometer-sized body with a speed comparable to the gravitational escape speed (Jewitt 2012).  Even particles launched too slowly to escape can be detached from Phaethon by solar radiation pressure while in flight, in a process called ``radiation pressure sweeping'' (Jewitt 2012).  At Phaethon's perihelion,  particles with $a \lesssim$ 0.25 mm (Equation (16) of Jewitt 2012) can be removed by radiation pressure sweeping once contact forces with the surface have been broken. On the day side the anti-solar direction of radiation pressure acceleration tends to push particles back into the surface but, around the terminator, radiation pressure sweeping can lead to escape.   The size of the largest particle that can be removed by radiation pressure sweeping scales as $r_H^{-2}$.  Even at $r_H$ = 1 AU, particles smaller than 5 $\mu$m can be expelled.   Selective loss of small particles from Phaethon is suggested by polarization studies and could result from this cause (Ito et al.~2018), although the interpretation is not unique (Shinnaka et al.~2018).  

The main problem for an origin of the Geminids by thermal fracture and/or desiccation stresses is that of rates.  The particles observed at perihelion are of micron size and the perihelion mass loss rate in these tiny grains is only $\sim$3 kg s$^{-1}$ (Jewitt and Li 2010, Li and Jewitt 2013, Jewitt et al.~2013, Hui and Li 2017). Larger, potentially mass-dominant particles could be launched at perihelion, but the absence of useful data makes the actual mass production rate very difficult to estimate.  In addition, observations show that the mass loss is restricted to a few days around perihelion.  Again, it is unclear how much of this restricted range is influenced by observational effects (principally phase darkening, which tends to make particles fade quickly as Phaethon swings around the Sun at perihelion).  Another problem for this hypothesis is the large particle launch speeds inferred for the Geminids.   Ryabova (2016) modeled the ejection speeds using the apparent width of the Geminid stream at 1 AU and found speeds $\sim$1 km s$^{-1}$, orders of magnitude too large to be produced by rock fracture.  Finally,  very large Geminids  (e.g.~Szalay et al.~2018 inferred a flux of 2 cm sized objects) probably cannot be launched by fracture.

\textit{Impact}:
A hypervelocity impact between Phaethon and a smaller asteroid  would naturally produce at least some debris with large launch speeds like those inferred by Ryabova (2016).  However, as noted above, the Geminid stream mass is about 10\% of the mass of Phaethon, and the Geminids were produced in the last $\sim 10^3$ years. For Phaethon to have lost 10\% of its mass within the last $\sim$10$^3$ years would imply an improbably short collisional lifetime of only a few $\times10^4$ years.  By comparison, Phaethon-sized asteroids in the denser (i.e.~more collision-prone) environment of the main-belt have collisional lifetimes $\gtrsim$10$^9$ years (Farinella et al.~1998, Bottke et al.~2005).  The latter is $\sim$10$^6$ times  longer than the Geminid stream age and 10$^2$ times longer even than the $\sim$26 Myr dynamical lifetime of Phaethon (de Le{\'o}n et al.~2010). A recent impact origin of the Geminids thus seems highly  improbable.  

\textit{Rotational Instability}:
Phaethon has a rotation period $\sim$3.6 hours (Ansdell et al.~2014, Hanus et al.~2016).  This is close to the reported rotational barrier period for C-type asteroids ($\sim$3.5 hours, Carbognani 2017), so that it is conceivable that rotational instability plays a role in the mass loss.  Two examples  of on-going rotational instability  have been identified in the main asteroid belt.  Mass loss from 311P/(2013 P5) (Jewitt et al.~2015b, 2018b) has been interpreted as ``surface shedding'' in which the weak, particulate outer layer of an asteroid is being rotationally cast off (Hirabayashi et al.~2015). However, the ejected mass is very small (nine pulses each of $\sim$10$^5$ kg were observed) from a body with mass $\sim 10^{11}$ kg; Jewitt et al.~2018b). The resulting mass ratio, $\sim 10^{-5}$, is tiny compared with the Geminid/Phaethon ratio $\sim$0.1.  Main-belt object P/2013 R3 experienced a more profound disruption, breaking   into about a dozen 100 m to 200 m scale fragments as a result of a presumed rotational instability brought about either by radiation or  mass-loss torques (Jewitt et al.~2017).  In both objects, the measured velocity dispersions were  $<$1 m s$^{-1}$, about 10$^3$ times smaller than the Geminid launch speed reported  by Ryabova (2016).   Thus, neither of the two best-characterized rotationally unstable asteroids presents a particularly compelling analogue for Phaethon, unless the reported high ejection speeds are in error.  We note that shape determinations and radar images (Taylor et al.~2019) are consistent with the presence of an equatorial skirt, perhaps produced by equatorward migration of surface material in response to centripetal acceleration.

\textit{Sublimation of Ice}:
The surface of Phaethon is too hot for exposed ice to exist but buried ice could, in principle,  survive.  The time taken for heat deposited on the surface to conduct to depth, $d$, is  $t_c \sim d^2/\kappa$, where $\kappa$ is the thermal diffusivity.  The largest plausible diffusivity is $\kappa \sim$ 10$^{-6}$ m$^2$ s$^{-1}$, appropriate for a compact dielectric solid (e.g.~rock).  Observations show that near-Sun mass loss is correlated with the time of perihelion to within $\sim$1 day (i.e.~$t_c \sim 10^5$ s), in which time heat can conduct to a characteristic depth $d \lesssim 0.3$ m.  However, the  temperature at 0.3 m depth ($T_{SS}/e \sim$ 370 K) is far too high for ice to exist there.  

Could more deeply-buried ice exist?  The presence of deeply buried ice cannot be reliably determined through calculation, since its long term stability depends on many unknowns in addition to the thermal diffusivity, including the permeability to gas flow, the initial abundance and spatial distribution of the ice (e.g.~single block vs.~separated ice chunks) and also on the poorly constrained orbital history of Phaethon (c.f.~Schorghofer and Hsieh 2018).  We simply note that the conduction time corresponding to the full radius of Phaethon (3 km) is $t_c \sim $ 0.3 Myr, again assuming $\kappa \sim$ 10$^{-6}$ m$^2$ s$^{-1}$.  On longer timescales, the core temperature would approach $\sim$300 K, the orbit-averaged temperature of Phaethon, and ice would again be unstable to sublimation.  The conduction time could be extended to match the 26 Myr dynamical lifetime by assuming much smaller values of the thermal diffusivity (specifically, $\kappa < 10^{-8}$ m$^2$ s$^{-1}$, Jewitt et al.~2018a), or by assuming that Phaethon was trapped into its present orbit much more recently  than 26 Myr ago (c.f.~Yu et al.~2019).

\clearpage

\section{SUMMARY}

We present Very Large Telescope observations of active asteroid (and Geminid parent) 3200 Phaethon  taken at closest approach to Earth (0.07 AU) at 10.7 $\mu$m wavelength.  The angular resolution of 0.3\arcsec~corresponds to 16 km at the distance of Phaethon. 

\begin{enumerate}

\item No extended emission attributable to dust is apparent in the VLT observations. We use the data and simple models to set limits to the presence of dust near Phaethon.  

\item An unbound (comet-like) coma would be detected if its optical depth exceeded $\sim 10^{-5}$, corresponding to a mass loss rate in submillimeter-sized particles $\sim$14 kg s$^{-1}$.  This is $\sim$50 times too small to allow the Geminids to be supplied by Phaethon in steady-state, which requires $\sim$700 kg s$^{-1}$.

\item The Hill sphere of Phaethon appears empty (optical depth $\lesssim 7\times10^{-5}$), consistent with the expectation that particles smaller than $\sim$10 cm are cleared by radiation pressure at each perihelion passage.

\item No quasi-linear dust trail is detected.  The corresponding upper limit to the optical depth is $\sim 6\times10^{-6}$.

\item No co-moving point sources (ejected secondary fragments) are detected, down to a size limit that varies strongly with projected angular distance from Phaethon but approaches 30 m radius at $\sim$150 km from Phaethon.

\item The observations are consistent with the complete inactivity of Phaethon when at 1 AU, and indicate that the production of the Geminids must occur episodically, through a process as-yet undetermined.

\end{enumerate}

\acknowledgments
We thank Yoonyoung Kim, Man-To Hui, Toshi Kasuga, Pedro Lacerda and the anonymous referee for comments on the manuscript.  Based on observations collected at the European Southern Observatory under ESO programmes 0100.C-0343, 098.C-0050 and 60.A-9800(I).  DA acknowledges support from the European UnionÕs Horizon 2020 Innovation program under the Marie Sklodowska-Curie grant agreement no. 793499 (DUSTDEVILS). This research made use of Astropy, a community-developed core Python package for Astronomy (Astropy Collaboration 2013).



{\it Facilities:}  \facility{VLT}.

\clearpage


\clearpage

\clearpage

\clearpage

\clearpage

\clearpage

\end{document}